\documentclass[12pt]{article}

\newcommand{\eps}{\varepsilon}

\newcommand{\old}[1]{{}}

\renewcommand{\equiv}{:=}

\newtheorem{theorem}{Theorem}
\newtheorem{lemma}{Lemma}

\begin{document}

\title{EINSTEIN'S EQUATIONS WITH ASYMPTOTICALLY STABLE CONSTRAINT
PROPAGATION~\footnote{Work supported by CONICOR, CONICET, the Swiss
  NSF, and SeCyT, UNC}}

\author{%
{\sc Othmar Brodbeck}
  \thanks{email: brodbeck@phys.psu.edu}\\
  {\small Center for Gravitational Physics and Geometry}\\
  {\small Department of Physics, The Pennsylvania State University}\\
  {\small University Park, PA 16802, USA}\\
  {\small and} \\ 
  {\small Institute for Theoretical Physics, The University of Zurich} \\
  {\small Winterthurerstrasse 190, 8057 Zurich, Switzerland} \\
 {\sc Simonetta Frittelli} 
  \thanks{email: simo@mayu.physics.duq.edu } \\
  {\small Duquesne University} \\
  {\small Pittsburgh, Pennsylvania, 15282, USA} \\
 {\sc Peter H\"ubner} 
  \thanks{email: pth@aei-potsdam.mpg.de} \\
  {\small Albert-Einstein-Institut} \\
  {\small Max-Planck-Institut f\"ur Gravitationsphysik} \\
  {\small Schlaatzweg 1, 14473 Potsdam, Germany} \\
 {\sc Oscar A. Reula}
  \thanks{Member of CONICET, email: reula@aei-potsdam.mpg.de}\\
  {\small FaMAF, Medina Allende y Haya de la Torre,}\\
  {\small Ciudad Universitaria, 5000 C\'ordoba, Argentina}\\
  {\small and} \\
  {\small Albert-Einstein-Institut} \\
  {\small Max-Planck-Institut f\"ur Gravitationsphysik} \\
  {\small Schlaatzweg 1, 14473 Potsdam, Germany} \\
}

\maketitle 

\vspace{-.4in}

\begin{abstract}
  We introduce a proposal to modify Einstein's equations by embedding them
  in a larger symmetric hyperbolic system.  The additional dynamical
  variables of the modified system are essentially first integrals of the
  original constraints.  The extended system of equations reproduces the
  usual dynamics on the constraint surface of general relativity, and
  therefore naturally includes the solutions to Einstein gravity. The
  main feature of this extended system is that, at least for a
  linearized version of it, the constraint surface is an attractor of
  the time evolution. This feature suggests that this system may be a
  useful alternative to Einstein's equations when obtaining numerical
  solutions to full, non-linear gravity.
\end{abstract}

\section{Introduction}
\label{sec:intro}
Over the past decade, computer power has increased to the point that
simulations of two- and even three-dimensional general relativity are now
feasible.  
These simulations, which assume little or no symmetry of their generic
field configurations, at first seemed to represent straightforward
generalizations of simpler one-dimensional calculations.
However, attempts to perform the higher dimensional simulations have
revealed a variety of unexpected features which limit accurate
simulations to a rather short time interval.
One such feature, which is believed to be a major source of numerical
error, is that numerical time evolution generates a rapidly growing
violation of the constraint equations.  
In this paper, we propose a system of dynamical equations
wherein the evolution naturally remains close to the constraint
surface. 
Although the most obvious application of this approach is to numerical
simulations, it may prove useful in other branches of general
relativity as well.

As is well known analytically, the time evolution predicted by the exact
Einstein equations is such that the constraint equations are satisfied on
each time slice when they are satisfied by the initial data.
Geometrically, the evolution vector field is tangential to the constraint
submanifold, implying that solutions to the complete set of equations are
insensitive to properties of the evolution field in the vicinity of
the constraint surface. 
In discrete approximations, on the other hand, the notion of tangency
is approximate, as is that of the constraint surface itself. 
As a consequence, the numerical evolution becomes sensitive to
possible instabilities of the constraint submanifold and numerical
solutions are, in general, carried away from it exponentially fast
with time.
Even in case one were able to construct a code whose
discretized vector field were exactly tangent to a discretised version of
the constraint submanifold, the same problem would be likely to arise,
as numerical errors on the initial data would prevent a start of the
time integration exactly on the constraint submanifold.

As demonstrated in \cite{Ch91co}, evolution schemes can be constructed in such a
way that the violation of the constraints has the same convergence order
as the scheme itself. 
This property, which is in the meantime a standard requirement for
evolution schemes, implies that the choice of an appropriately fine grid
is sufficient to satisfy the constraint equations at any given time
with arbitrary accuracy.  
However, since the violation of the constraints grows very quickly
with time, the utility of grid refinements to reduce constraint
violations is very limited, especially in two- and three-dimensional
calculations.

In the so-called constrained evolution schemes one, attempts to solve
this problem by isolating two sets of variables in Einstein's
equations. One uses evolution equations to evolve one set and
determines the variables of the other set by solving constraint
equations on each time slice.
This method has frequently been used in one-dimensional simulations
where, on the one hand, it is easy to split the variables into
dynamical and longitudinal ones, and where, on the other hand, the
constraint equations are ordinary differential equations along a
space-like direction.
However, in two- and three-dimensional simulations with space-like
hypersurfaces as time surfaces, the elliptic character of the
constraint equations makes it expensive in computer time to solve the
constraint equations on each time slice.\footnote{Not to mention the
  problems arising in the treatment of the grid boundaries.}
Furthermore, this approach does not guarantee that the complete
set of Einstein's equations is solved. 
Since only a subset of the variables is determined by evolution
equations, some of these equations remain unused. 
The problem is, therefore, shifted to the preservation of the unused
evolution equations, which, as shown in~\cite{De87eo}, is a problem of
similar nature.

In order to guarantee a good approximation to the complete set of
field equations, it is, therefore, necessary to analyze the behavior
of the evolution vector field in a whole neighborhood of the
constraint submanifold.
Away from the constraint submanifold the evolution field is not
uniquely determined as field configurations violating the constraint
equations are physically not relevant.
Hence, the evolution vector field can be modified in an arbitrary
way, as long as its values on the constraint submanifold
remain unchanged, and as long as the modified field continues to be
strongly hyperbolic, so that the Cauchy problem is well posed in a
whole neighborhood of the constraint surface.

Of particular interest are modified equations for which the constraint
submanifold is asymptotically stable, because for equations with this
feature, sufficiently accurate codes are expected to generate solutions
which remain close to the constraint surface, and which, therefore, would
represent improved approximations to Einstein's equations.

Modifications of the evolution vector field have previously been studied.
However, in general these preserve the time reversal symmetry of
Einstein's equations, which implies that modifications of this type
cannot have the desired properties.
Time reversal symmetry implies that if the evolution field is such
that a solution to some initial data in a neighborhood of the
constraint submanifold approaches the constraint submanifold 
during time evolution, then the solution to the time reversed initial
data will asymptotically be carried away from this submanifold. 
Thus, without a modification of Einstein's equations which breaks the
time reversal symmetry, the best one can expect to achieve is a set of
equations for which the constraint submanifold is stable, but not
asymptotically stable.
However, stability of the undiscretized equations is not sufficient
for numerical simulations, since spurious solutions to the discretized
equations can grow very rapidly even for stable systems.
To eliminate the impact of such solutions, it is, therefore, necessary that
the constraint manifold is an attractor for the time evolution. 

The above-mentioned modifications of Einstein's equations are, in
general, obtained by including dynamical quantities which are proportional
to the constraint expressions. An alternative argument showing that
extensions of this type cannot lead to an asymptotically stable constraint
surface is the following:
Since the constraint equations are of the same order as the
evolution equations, their inclusion affects mainly the principal
part of the evolution equations, whence the freedom remaining after
requiring that these terms do not destroy strong hyperbolicity is very
limited.
Thus, such extensions ensure the well posedness of the problem, but
not the asymptotic stability. 
This can only be obtained either via modifications of the lower order
terms or the addition of higher (second) order terms, that is, by including
damping or diffusion terms.

In the next section, we propose a modification of Einstein's equations
which includes new dynamical terms proportional to certain first integrals
of the constraint expressions, rather than to the constraints themselves. 
The dissipation, that is the time asymmetry is not of the diffusive
type,\footnote{One could also introduce diffusive dissipation, but
  this would significantly reduce the allowed maximal time step in
  explicit discretisation schemes.} and is built into the definition
of these integrals.

We show that the Cauchy problem for the resulting new system, which we
call the {\bf $\lambda$-system},\footnote{The name is a remnant of the
  way the system was originally guessed by Brodbeck and H\"ubner,
  namely by a formal application of Lagrangian multiplier techniques.}
is locally well posed.
We also prove that if the constraints are initially satisfied,
and if their first integrals initially vanish, then the
$\lambda$-system provides solutions to Einstein's equations.  
Moreover, for initial data sets for the $\lambda$-system,
which are sufficiently close to the constraint submanifold and
sufficiently close to zero, respectively, we suspect that the
solutions asymptotically tend 
to solutions to Einstein's equations.

In section~\ref{sec:LG}, we give support to our expectation by proving
that the linearized extended system is asymptotically stable, thus
showing that in the linearized case, the constraint submanifold is
indeed an attractor for the $\lambda$-system.
In section~\ref{sec:conclu}, we discuss further expectations
in connection with our proposal.

\section{The \mbox{\boldmath$\lambda-$system}}
\label{sec:modeq}

In this section we spell out our proposal for a modification of
Einstein's equations with an asymptotically stable constraint
submanifold. 
For definiteness, we choose the symmetric
hyperbolic system introduced by Frittelli and Reula
in~\cite{FrittelliS:NewLG}, which corresponds to the parameters
$\alpha=\beta=-1$ in~\cite{FrittelliS:FirOS}. Although the full
equations (with the non-principal part terms added) are given
in~\cite{St98tc}, we repeat them for completeness.

In the version of Einstein's vacuum equations chosen, the
system is given by the following set of dynamical equations (where
Latin indices run from 1 to 3):
\begin{eqnarray}
   \dot{h}^{ij}                                                 &=&
          N^nh^{ij}{},_n
        + Q\sqrt{h}\left(2P^{ij}-Ph^{ij}\right)
        - 2h^{n(i}N^{j)}{},_n  \;  ,    \label{originalhdot}      \\
   \dot{M}^{ij}{}_k                                             &=&
          N^n M^{ij}{}_{k,n}
        {} + Q\sqrt{h}\left(P^{ij}{},_k-2\delta_k{}^{(i}P^{j)n}{},_n\right)
                                                        \nonumber\\
                                                                & &
        {} + Q\sqrt{h}\Bigg(
          \frac32 P^{ij}M_k-PM^{ij}{}_k+Q^{-1}P^{ij}Q,_k \nonumber\\
                                                                & &
        \quad {} - 2 \delta_k{}^{(i}\Big[
          h^{j)q} h_{mr}h_{ns}P^{mn}M^{rs}{}_q
         -2M^{j)p}{}_nP^{mn}h_{pm}                            \nonumber\\
                                                                & &
        \qquad {} +\frac32 P^{j)n}M_n
         -\frac12 h^{j)n}PM_n
                        \Big]
                   \Bigg)                                \nonumber\\
                                                                & &
        {} + h^{ij} N^n{},_{nk}
        - h^{n(i}N^{j)}{},_{nk}
        - 2N^{(i}{},_nM^{j)n}{}_k
        + N^m{},_kM^{ij}{}_m   \;     ,     \label{originalMdot}  \\
   \dot{P}^{ij}                                                 &=&
        N^nP^{ij}{},_n
        + Q\sqrt{h}\left(
        h^{mn}M^{ij}{}_{m,n}-2h^{n(i}M^{j)k}{}_{k,n}\right)\nonumber\\
                                                                & &
        {} + Q\sqrt{h}\Bigg(
        4h_{np}h^{m(i}M^{j)n}{}_kM^{kp}{}_m
        -h^{ik}h^{jn}h_{rp}h_{sq}M^{rs}{}_kM^{pq}{}_n   \nonumber\\
                                                                & &
        \quad {} +\frac12 h^{ik}h^{jn}M_kM_n
        +2M^{nk}{}_kM^{ij}{}_n
        -2M^{ik}{}_nM^{jn}{}_k                          \nonumber\\
                                                                & &
        \quad {} -2h_{mn}h^{kp}M^{im}{}_kM^{jn}{}_p
        -2h^{n(i}M^{j)k}{}_kM_n
        +M^{ij}{}_nh^{nk}M_k                            \nonumber\\
                                                                & &
        \quad {} -Q^{-1}\Big[
        h^{ik}h^{jn}Q,_{kn}
        +2M^{k(i}{}_nh^{j)n}Q,_k
        -M^{ij}{}_mh^{mk}Q,_k                           \nonumber\\
                                                                & &
        \qquad {} -h^{ij}\left(h^{kn}Q,_{kn}+2M^{km}{}_mQ,_k\right)
                \Big]
        +2P^{ik}h_{kn}P^{nj}
        -\frac32 P P^{ij}                               \nonumber\\
                                                                & &
        \quad {} +h^{ij}(\frac12P^2
        -h_{mr}h_{ns}P^{mn}P^{rs}
        \Big) \Bigg)  
        -2P^{k(i}N^{j)}{},_k \; .     \label{originalPdot}              
\end{eqnarray}
Here, $h^{ij}$ is the inverse intrinsic metric of the spacelike
hypersurfaces $\Sigma_t$, $P^{ij} := k^{ij} - h^{ij} k$ denotes a
linear combination of the extrinsic curvature $k^{ij}$ of the slice
and its trace $k$, and $M^{ij}{}_k := \frac{1}{2} \left( h^{ij}{}_{,k}
  - h^{ij} h_{rs} h^{rs}{}_{,k} \right)$
represents a linear combination of spatial derivatives of the 
inverse intrinsic metric.
The functions $P$ and $M_k$ are abbreviations for $h_{ij} P^{ij}$ and
$h_{ij} M^{ij}{}_k$, respectively, and $Q$ and $N^i$ are arbitrary
given functions fixing the gauge degrees of freedom.

This evolution system is symmetric hyperbolic with respect to the
inner product
\begin{eqnarray}
  \label{eq:He}
&&  \langle     h_1^{ij}, P_1^{ij}, M_1^{ij}{}_k \mid
     H_e \mid h_2^{ij}, P_2^{ij}, M_2^{ij}{}_k \rangle 
:= \nonumber\\
&& \int_{\Sigma_t} \left\{ 
         e_{im}e_{jn} \bar h_1^{ij} h_2^{mn} 
       + e_{im}e_{jn} \bar P_1^{ij} P_2^{mn} 
       + e_{im}e_{jn}e^{kl} \bar M_1^{ij}{}_k M_2^{mn}{}_l 
   \right\} \, d\Sigma \; ,
\end{eqnarray}
where $e_{ij}$ denotes an Euclidean flat metric on the hypersurface
$\Sigma_t$.
It is completed by the following set of constraints
equations:
\begin{equation}
   {\cal C} = 0 , \qquad
   {\cal C}^i = 0  , \qquad
   {\cal C}^{ij}{}_k =  0,
\end{equation}
where
\begin{eqnarray}
   {\cal C}                                     & \equiv &      
        -M^{kn}{}_{n,k} 
        +h_{pq}M^{pk}{}_nM^{qn}{}_k
        -M^{kq}{}_qM_k
        +\frac14 h^{kn}M_kM_n                           \nonumber\\
                                                &        &
        {} -\frac12 h_{mn}h_{rs}h^{pq}M^{mr}{}_pM^{ns}{}_q
        -\frac12 h_{mn}h_{rs}P^{mr}P^{ns}
        +\frac14 P^2 \; ,                              \label{scalar}\\
   {\cal C}^i                                    & \equiv & 
        P^{ik}{},_k
        -2h_{mn}M^{im}{}_kP^{nk}
        -\frac12 h^{ik}PM_k                             \nonumber\\
                                                &         &
        {} +h_{mn}h_{pq}h^{ik}M^{mp}{}_kP^{nq}
        +\frac32 P^{ik}M_k \; ,                        \label{vector}\\
   {\cal C}^{ij}{}_k                            & \equiv & 
        2M^{ij}{}_k 
        -h^{ij}h_{pq}M^{pq}{}_k
        -h^{ij}{},_k \; .                              \label{M}
\end{eqnarray}

The first two constraints are the scalar and the vector constraint of
Einstein's equations, that is the time-time and time-space components
of the Einstein tensor for a given 3+1 decomposition of space-time. 
The third is the statement that the tensor $M^{ij}{}_k$ is a
linear combination of spatial derivatives of the 3-metric.

To solve the initial value problem of general relativity in this
approach, one prescribes an initial data set $(h_0^{ij}, P_0^{ij},
M_0^{ij}{}_k)$ at $t=0$ which satisfies the constraints equations and
subsequently solves the above evolution equations.
Symmetric hyperbolicity of the evolution system implies that a unique
local solution does exist.
 
By taking a time derivative of equations (\ref{scalar}--\ref{M}) and
using (\ref{originalhdot}--\ref{originalPdot}) to eliminate time
derivatives in favor of spatial derivatives, the following evolution
equations for the constraints are obtained:
\begin{eqnarray}
   \dot{\cal C}                                         & = &
         N^n{\cal C},_n
        +3 Q\sqrt{h} {\cal C}^k{},_k + \, \ldots \, ,   \label{cdot} \\
   \dot{\cal C}^i                                       & = &   
         N^n{\cal C}^i,_n
        +Q\sqrt{h}\left(h^{ik}{\cal C},_k
        +h^{rs}{\cal C}^{ik}{}_{[r,k]s}
        +h^{is}h^{kl}h_{mn}{\cal C}^{mn}{}_{[s,k]l} \right) 
                                                       \nonumber\\
      &        &
        {} + \, \ldots \; ,                              \label{second}   \\
   \dot{\cal C}^{ij}{}_k                                        & = &   
         N^n{\cal C}^{ij}{}_{k,n}
        - 2Q\sqrt{h}\left(2\delta_k{}^{(i}{\cal C}^{j)} 
        - h^{ij} h_{kl} {\cal C}^l\right)                     
        + \, \ldots \; ,
                                                     \label{cijkdot}
\end{eqnarray}
where ``$\ldots$'' represent undifferentiated terms which are linear in the
constraint quantities and at least linear in the variables $P^{ij}$
and $M^{ij}{}_k$.
 
Since equation (\ref{second}) is of second order in spatial
derivatives, we introduce a further constraint by\footnote{One could
  also consider the constraint $C^i{}_j \equiv C^{ik}{}_{jk}$, which
  still makes the constraint system symmetric hyperbolic and produces
  a smaller number of extra fields.}
\begin{equation}
\label{cijkl}
  C^{ij}{}_{kl}:= 2M^{ij}{}_{[k,l]} + 2M^{ij}{}_{[k}M_{l]}.
\end{equation}
By taking a time derivative of (\ref{cijkl}), we obtain
\begin{equation}
   \dot{\cal C}^{ij}{}_{kl}                                     =  
         N^n{\cal C}^{ij}{}_{kl,n}
        -2Q\sqrt{h}\left( \delta_k{}^{(i}{\cal C}^{j)}{},_l
        -\delta_l{}^{(i}{\cal C}^{j)}{},_k\right)
        + \, \ldots \; ,                                \label{cijkldot}
\end{equation}
and by plugging (\ref{cijkl}) into (\ref{second}), we see that the
evolution equation for ${\cal C}^i$ can be rewritten as
\begin{equation}
   \dot{\cal C}^i                                        =
         N^n{\cal C}^i,_n
        +Q\sqrt{h}\left(h^{ik}{\cal C},_k
        +h^{rs}{\cal C}^{ik}{}_{rk,s}\right)
        + \, \ldots  \; .                               \label{cidot}        
\end{equation}

The constraint quantities ${\cal C}$, ${\cal C}^i$, ${\cal C}^{ij}{}_k$,
and  ${\cal C}^{ij}{}_{kl}$ thus propagate according to the first-order
system of equations consisting of (\ref{cdot}), (\ref{cijkdot}), and
(\ref{cijkldot}--\ref{cidot}), which is  symmetric hyperbolic with
respect to the following inner product: 
\begin{eqnarray}
  \label{eq:Hc}
&& \langle C_1, C_1^i, C_1^{ij}{}_{k}, C_1^{ij}{}_{kl} \mid
           H_C \mid C_2, C_2^i, C_2^{ij}{}_{k}, C_2^{ij}{}_{kl} \rangle 
:= \nonumber\\
&&\qquad {} \int_{\Sigma_t} \Big\{
        \frac13        \bar C_1 C_2 
     +          e_{ij} \bar C_1^i C_2^j 
     +          e_{ij} e_{kl} e^{rs} \bar C_1^{ik}{}_{r} C_2^{jl}{}_{s}
 \nonumber\\
 & & \qquad \qquad {} + \frac14 e_{im} e_{jn} e^{kp} e^{lq}
                   \bar C_1^{ij}{}_{kl} C_2^{mn}{}_{pq}
                   \Big\}\,d\Sigma \; .
\end{eqnarray}
Uniqueness of solutions to this system implies that if the constraints 
are initially satisfied, then the exact evolution equations preserve
them.
When, as in numerical simulations, the constraint variables
initially are not precisely zero, then the corresponding solution is,
in general, carried away from the constraint surface during time
evolution. 
However, since the evolution equations for the constraint variables are
symmetric hyperbolic, the violation of the constraints becomes smaller
when the constraints initially are satisfied with better
accuracy.

In order to obtain a system with an asymptotically stable constraint
submanifold, we propose a modification of Einstein's equations, which
is inspired by the behavior of dissipative systems, where a transient
eventually is dissipated away as the system settles down.
We extend the set of dynamical variables by considering the following
``time integrals'' of the constraint variables:
\begin{eqnarray}
  \label{eq:lambda1}
    \dot \lambda 
                &=& \alpha_0 {\cal C} - \beta_0 \lambda \; , \\
  \label{eq:lambdai1}
    \dot \lambda^i 
                &=& \alpha_1 {\cal C}^i - \beta_1 \lambda^i \; , \\ 
  \label{eq:lambdaijk1}
    \dot \lambda^{ij}{}_k 
                &=& \alpha_3 {\cal C}^{ij}{}_k - \beta_3 \lambda^{ij}{}_k \; , \\ 
  \label{eq:lambdaijkl1}
    \dot \lambda^{ij}{}_{kl} 
                &=& \alpha_4 {\cal C}^{ij}{}_{kl} - \beta_4
                \lambda^{ij}{}_{kl} \; ,  
\end{eqnarray}
where the tensor-valued $\lambda$-variables are assumed to have the
same symmetries as the corresponding $C$-variables, and where
$\alpha_i \neq 0$ and $\beta_i>0$ are constants.

The equations (\ref{eq:lambda1}--\ref{eq:lambdaijkl1}) represent
evolution equations for the $\lambda$-variables which in terms of the
fundamental variables $(h^{ij}, P^{ij}, M^{ij}{}_k)$ are given by 
\begin{eqnarray}
  \label{eq:lambda2}
         \dot \lambda &=& \alpha_0\left(-M^{kn}{}_{n,k}        
        +h_{pq}M^{pk}{}_nM^{qn}{}_k
        -M^{kq}{}_qM_k
        +\frac14 h^{kn}M_kM_n\right)  \nonumber\\
                                                        & &
        {} - \beta_0 \lambda \; ,\\
  \label{eq:lambdai2}
         \dot \lambda^i &=& \alpha_1 \left(P^{ik}{},_k
        -2h_{mn}M^{im}{}_kP^{nk}
        -\frac12 h^{ik}PM_k\right)     - \beta_1 \lambda^i \; ,\\ 
  \label{eq:lambdaijk2}
         \dot \lambda^{ij}{}_k &=&
         \alpha_3 \left( 2M^{ij}{}_k - h^{ij}h_{rs}M^{rs}{}_k
         - h^{ij}{},_k \right)
         - \beta_3 \lambda^{ij}{}_k \; ,\\ 
  \label{eq:lambdaijkl2}
         \dot \lambda^{ij}{}_{kl} &=& 
         \alpha_4 \left( 2M^{ij}{}_{[k,l]} + 2M^{ij}{}_{[k}M_{l]}\right) 
         - \beta_4 \lambda^{ij}{}_{kl} \; .
\end{eqnarray}
In the present form, the combined system
(\ref{originalhdot}--\ref{originalPdot},\ref{eq:lambda2}--\ref{eq:lambdaijkl2})
is not symmetric hyperbolic, since the equations
(\ref{eq:lambda2}--\ref{eq:lambdaijkl2}) involve spatial derivatives of
the variables $(h^{ij}, P^{ij}, M^{ij}{}_k)$, whereas
the equations (\ref{originalhdot}--\ref{originalPdot}) do not contain
$\lambda$-variables at all.
However, by adding terms containing first derivatives of the
$\lambda$-variables it is possible to bring the system
(\ref{originalhdot}--\ref{originalPdot},\ref{eq:lambda2}--\ref{eq:lambdaijkl2}) 
into a symmetric hyperbolic form,
\begin{eqnarray}
  \dot{h}^{ij}                                                 &=&
        \alpha_3 h^{mn}\lambda^{ij}{}_{m,n}
        + N^nh^{ij}{},_n
        + \mbox{ source terms} \; ,             \label{modhdot}          \\
   \dot{M}^{ij}{}_k                                             &=&
        2 \alpha_4 h^{lm}\lambda^{ij}{}_{kl,m}
        -\alpha_0 \delta_k{}^{(i}h^{j)l}\lambda,_l                 \nonumber\\
                                                                & &
        {} + N^n M^{ij}{}_{k,n}
        + Q\sqrt{h}\left(P^{ij}{},_k-2\delta_k{}^{(i}P^{j)n}{},_n\right)
                                                        \nonumber   \\
                                                                & &
        {} + \mbox{ source terms} \; ,         \label{modMdot}          \\
   \dot{P}^{ij}                                                 &=&
         \alpha_2 h^{l(i}\lambda^{j)}{},_l                \nonumber \\
                                                                & &
        {} + N^nP^{ij}{},_n
        + Q\sqrt{h}\left(
        h^{mn}M^{ij}{}_{m,n}-2h^{n(i}M^{j)k}{}_{k,n}\right)\nonumber\\
                                                                & &
        {} + \mbox{ source terms} \; .           \label{modPdot}       
\end{eqnarray}
By construction, the ``$\lambda$-system''
(\ref{eq:lambda2}--\ref{modPdot}) is symmetric hyperbolic with respect 
to the inner product
\begin{eqnarray}
  \label{eq:bigHe}
&&  \langle  h_1^{ij}, P_1^{ij},M_1^{ij}{}_k,
             \lambda_1,\lambda^i_1,\lambda_1^{ij}{}_k,\lambda_1^{ij}{}_{kl}
             \mid H_E^{\lambda} \mid
             h_2^{ij}, P_2^{ij}, M_2^{ij}{}_k,
             \lambda_2,\lambda^i_2,\lambda_2^{ij}{}_k,\lambda_2^{ij}{}_{kl} \rangle 
:= \nonumber\\
&& \qquad \int_{\Sigma_t} \Big\{ 
          e_{im}e_{jn} \bar h_1^{ij} h_2^{mn} 
        + e_{im}e_{jn} \bar P_1^{ij} P_2^{mn} 
        + e_{im}e_{jn}e^{kl} \bar M_1^{ij}{}_k M_2^{mn}{}_l 
        + \bar\lambda_1 \lambda_2
        \nonumber\\
                                                        &&
   \qquad \qquad {}
        + e_{ij} \bar\lambda_1^i \lambda_2^j
        + e_{ip}e_{jq}e^{kr}\bar\lambda_1^{ij}{}_{k}
                                \lambda_2^{pq}{}_{r}
        + e_{ip}e_{jq}e^{kr}e^{ls}\bar\lambda_1^{ij}{}_{kl}
                                      \lambda_2^{pq}{}_{rs}
   \Big\} \, d\Sigma \; . 
\end{eqnarray}

The initial data for this purely dynamical set of equations consists
of arbitrary functions
\begin{equation}
\label{fulldataset}
(h_0^{ij}, P_0^{ij}, M_0^{ij}{}_k, \lambda_0, \lambda^i_0,
\lambda_0^{ij}{}_k,\lambda^{ij}_0{}_{kl}) \; .
\end{equation}
However, the dynamical degrees of freedom are extended by 40
$\lambda$-variables.

Clearly, for an arbitrary solution to Einstein's
equations, $(h_E^{ij}, P_E^{ij}, M_E^{ij}{}_k)$,
the embedded field configuration $(h^{ij}, P^{ij},\allowbreak
M^{ij}{}_k,\lambda, \allowbreak \lambda^i,
\lambda^{ij}{}_{k},\allowbreak \lambda^{ij}{}_{kl}) :=  (h_E^{ij},
P_E^{ij},\allowbreak M_E^{ij}{}_k,0,0,\allowbreak 0,0)$ is 
a solution to the $\lambda$-system.
Conversely, every solution to the $\lambda$-system with vanishing
$\lambda$-variables is also a solution to Einstein's
equations.
Due to this property, and since the solutions to the $\lambda$-system
are unique, the $\lambda$-system naturally reproduces the 
dynamics on the constraint submanifold of general relativity. 

Note that if the constraints are initially not satisfied, then, even when
the $\lambda$-variables initially vanish, the $\lambda$-variables 
would pick up a non-zero value during time evolution. 
Hence, solutions to the $\lambda$-system corresponding to such initial
data sets would not represent solutions to the complete set of
Einstein's equations.
In fact, they would not even solve the evolution equations of general
relativity. 
However, for constraint- and $\lambda$-variables which initially are
sufficiently close to zero, we suspect that the solutions
asymptotically approach solutions to the Einstein equations. 
In the following section, we give analytical evidence that this
conjecture could be true.

The system is by no means uniquely ``extended'', since one could still
add non-principal (undifferentiated) terms, as long as they vanish when
$\lambda=\lambda^i=\lambda^{ij}{}_{k}=\lambda^{ij}{}_{kl}=0$.
Such terms might be useful in order to treat the strongly non-linear
regime.
Of particular interest might be to choose the coefficients $\alpha_i$
and $\beta_i$, which control the damping in the $\lambda$-equations,
to be quadratic functions of the basic variables
$(h^{ij},P^{ij},M^{ij}{}_k)$, so that the damping becomes stronger at 
points where the non-linearities intensify.

It is fairly easy to implement similar schemes for alternative symmetric
hyperbolic systems for the Einstein equations, as well as for symmetric
hyperbolic systems for other theories with constraints, like, for
instance, Yang--Mills theories.
The strategy is the same: One writes equations with damping for first
integrals of the constraints and modifies the evolution equations
such that the extended system becomes symmetric hyperbolic.
This can always be achieved, because the inclusion of the new equations
modifies an off diagonal sector of the principal symbol matrix.

\section{Asymptotic stability of the constraint propagation}
\label{sec:LG}

The inclusion of the $\lambda$-terms into
(\ref{originalhdot}--\ref{originalPdot}) affects, in turn, the
evolution of the constraint quantities ${\cal C}, {\cal C}^i$, ${\cal
  C}^{ij}{}_{k}$, and ${\cal C}^{ij}{}_{kl}$. 
Recalculating the time derivative of these, and using
(\ref{modhdot}--\ref{modPdot}), yields the constraint evolution
equations for the new system,
\begin{eqnarray}
  \dot{\cal C}                                          & = &
         N^n{\cal C},_n
        +3 Q\sqrt{h} {\cal C}^k{},_k 
        - 2\alpha_4 h^{mn}\lambda^{kl}{}_{km,nl}
        + 2\alpha_0 h^{mn}\lambda,_{mn}
        + \, \ldots \; ,                                \label{clambda}\\
   \dot{\cal C}^i                                       & = &   
         N^n{\cal C}^i,_n
        +Q\sqrt{h}\left(h^{ik}{\cal C},_k
        +h^{rs}{\cal C}^{ik}{}_{rk,s}\right)
        +\alpha_1 h^{m(n}\lambda^{i)},_{mn}    
        + \, \ldots \; ,                                \label{cilambda}\\
   \dot{\cal C}^{ij}{}_{k}                             & = &   
         N^n{\cal C}^{ij}{}_{k,n}
        - 2Q\sqrt{h}\left(2\delta_k{}^{(i}{\cal C}^{j)}- h^{ij}h_{kl}C^l
                                          \right)               \nonumber\\
                                                        &    &
        {} + 2\alpha_3 h^{mn}\lambda^{ij}{}_{m,nk} 
        + 2\alpha_4 h^{mn}\left(2\lambda^{ij}{}_{km},{}_n 
        - h^{ij}h_{rs}\lambda^{rs}{}_{km},{}_n\right)           \nonumber\\
                                                        &    &
        {} - \alpha_0 \left( 2\delta^{(i}{}_k h^{j)l}\lambda{},{}_l 
        - h^{ij}\lambda{},{}_k \right)
        + \, \ldots \; ,                                \label{cijklambda} 
        \\
   \dot{\cal C}^{ij}{}_{kl}                             & = &   
         N^n{\cal C}^{ij}{}_{kl,n}
        -2Q\sqrt{h}\left( \delta_k{}^{(i}{\cal C}^{j)}{},_l
        -\delta_l{}^{(i}{\cal C}^{j)}{},_k\right)               \nonumber\\
                                                        &    &
        {} + 2\alpha_4 \left(h^{mn}\lambda^{ij}{}_{km,nl}
        - h^{mn}\lambda^{ij}{}_{lm,nk}\right)                       \nonumber\\
                                                        &    &
        {} - \alpha_0 \left(\delta_k{}^{(i}h^{j)m}\lambda,_{ml}
        +\delta_l{}^{(i}h^{j)m}\lambda,_{mk}\right)
        + \, \ldots \; .                                \label{cijkllambda}
\end{eqnarray}
Again ``$\ldots$'' represent undifferentiated terms that are linear in the
constraint quantities and at least linear in $(P^{ij}, M^{ij}{}_k)$.

The propagation of the constraints is ruled by the system of equations
consisting of (\ref{eq:lambda1}--\ref{eq:lambdaijkl1}) and
(\ref{clambda}--\ref{cijkllambda}), which determines whether or not the
constraints asymptotically ``decay'' to zero.
The crucial feature of this system is that the right hand side
also contains non-principal terms.
Roughly speaking, the operator on the right-hand side amplifies
constraint violations if the matrix representing its action on periodic
functions has any eigenvalue with a positive real part.  
On the other hand, if all the eigenvalues have a negative
real part, the operator induces an asymptotic decay of these violations.

Instead of attacking the full non-linear problem as stated, which
represents a problem well beyond the scope of present analytical
techniques, we consider the linear regime of general relativity.
That is, we restrict attention to 3-metrics of the form $h^{ij}=
e^{ij} + \epsilon \gamma^{ij}$ with $e^{ij}=\delta^{ij}$ and
$\epsilon\ll 1$.
This implies that the variables $(P^{ij}, M^{ij}{}_k)$ are of
first-order in $\epsilon$, as are the constraint quantities $({\cal
  C},{\cal C}^i, {\cal C}^{ij}{}_k, {\cal C}^{ij}{}_{kl})$ and the
variables $(\lambda,\lambda^i, \lambda^{ij}{}_k, \lambda^{ij}{}_{kl})$. 
Thus, the terms represented by ``$\ldots$'' in the
equations~(\ref{clambda}--\ref{cijkllambda}) are of second order in
$\epsilon$ and shall be neglected.
Without loss of generality, we restrict the following arguments to the
case where the gauge source functions $Q$ and $N^i$ are constant.  
All arguments that follow refer to this linearized regime.  

Although we lack a proof for the non-linear case, the following
considerations provide analytical evidence for the asymptotic
stability of the constraint propagation, in particular since the full
evolution equations are quasi-linear.

For, as we believe, purely technical reasons, we adopt the following
choice of coefficients: 
$\beta_0 = \beta_1 = \beta_3 = \beta_4 \equiv \beta>0$ and 
$\alpha_4 = \frac{\sqrt{3}}2 \alpha_0$.
\begin{theorem}
With the above assumptions, the constraint submanifold
of the linearized Einstein equations is an asymptotically 
stable submanifold for the solutions to the linearized, $\lambda$-extended
Einstein equations.
\end{theorem}

We partition the proof of this theorem in several lemmas:
We first show that the initial value problem is well posed and that the
solutions stay bounded with time.
Thus, it is possible to apply Laplace transformation techniques, which
reduce the problem to the study of the eigenfrequencies of the system.
For these frequencies, we show that the real part is non-positive, only
approaches zero as the wave number goes to zero, and does so
quadratically.
Then stability follows from estimates in \cite{KreissG:StabCL}.

Without loss of generality, we expand
the linearized dynamical fields in 
Fourier integrals of the the following form:
\begin{eqnarray}
    \lambda\,(x,t)&=&
    \int\hat\lambda\,
    (k,t) \exp(ik\!\cdot\! x)\; d^3k\;\,,\\
    \lambda^i\,(x,t)&=&
    \int\hat\lambda^i\,
    (k,t) \exp(ik\!\cdot\! x)\; d^3k\;\,,\\
     &\vdots&\\
    C^{ij}{}_{k}\,(x,t)&=&
    \int\hat C^{ij}{}_{k}\,
    (k,t) \exp(ik\!\cdot\! x)\; d^3k\;\,,\\
    C^{ij}{}_{kl}\,(x,t)&=&
    \int\hat C^{ij}{}_{kl}\,
    (k,t) \exp(ik\!\cdot\! x)\; d^3k\;\,,
\end{eqnarray}
where $k\!\cdot\! x:= k_i x^i$. 

In terms of the Fourier transformed variables,
equation~(\ref{clambda}--\ref{cijkllambda}) and (\ref{eq:lambda1}--
\ref{eq:lambdaijkl1}) reduce to the system of ordinary differential
equations given by  
\begin{eqnarray}
\label{fourierlinpmlambda}
   \dot {\hat{\lambda}}                         &=& 
        - \beta \hat{\lambda}  
        +\alpha_0 \hat{\cal C}\;,     
                                              \label{lambda0}\\
   \dot {\hat{\lambda}}{}^i                       &=& 
        - \beta \hat{\lambda}^i
        + \alpha_1 \hat{\cal C}^i\;,
                                              \label{lambda1}\\ 
   \dot {\hat{\lambda}}{}^{ij}{}_{kl}             &=& 
        - \beta \hat{\lambda}^{ij}{}_{kl}
        +\frac{\sqrt{3}}2 \alpha_0 \hat{\cal C}^{ij}{}_{kl}\;,
                                              \label{lambda4}\\
   \dot{\hat{\cal C}}                                   & = &
        ik_n N^n\hat{\cal C}
        +3 iQk_m \hat{\cal C}^m{}
        + \sqrt{3} \alpha_0 \hat{\lambda}^{rl}{}_{rm} k^m k_l
        - 2\alpha_0 \hat{\lambda} k^n k_n\;,     \label{c0}\\
   \dot{\hat{\cal C}}{}^i                                 & = &   
         ik_nN^n\hat{\cal C}^i
        +iQ\left(k^i\hat{\cal C}
        +k^r\hat{\cal C}^{in}{}_{rn}\right)
        - \frac12\alpha_1 (k^nk_n\hat{\lambda}^i   
        + k^ik_n\hat{\lambda}^n)\;,             \label{c1}\\   
   \dot{\hat{\cal C}}{}^{ij}{}_{kl}                       & = &   
        i k_n N^n \hat{\cal C}^{ij}{}_{kl}
        - 2 i Q \left( \delta_k{}^{(i} \hat{\cal C}^{j)} k_l
        -\delta_l{}^{(i} \hat{\cal C}^{j)} k_k \right)    \nonumber\\
                                                       &    &
        {} - \sqrt{3} \alpha_0 \left(\hat{\lambda}^{ij}{}_{kr} k^r k_l
        - \hat{\lambda}^{ij}{}_{lr} k^r k_k\right)          \nonumber\\
                                                       &    &
        {} + \alpha_0 \hat{\lambda}
           \left( \delta_k{}^{(i} k^{j)} k_l
                  - \delta_l{}^{(i} k^{j)} k_k \right)\;,                    \label{c4}
\end{eqnarray}
and
\begin{eqnarray}
   \dot {\hat{\lambda}}{}^{ij}{}_{k}             &=& 
        - \beta \hat{\lambda}^{ij}{}_{k} 
        + \alpha_3 \hat{\cal C}^{ij}{}_{k}\;,
                                                \label{lambda3}\\
   \dot{\hat{\cal C}}{}^{ij}{}_{k}                       & = &   
        ik_n N^n\hat{\cal C}^{ij}{}_{k}
         -\alpha_3\hat{\lambda}^{ij}{}_{m}k^mk_k + 
          \hat S^{ij}{}_k
         \label{c3}\;\,,
\end{eqnarray}
where
\begin{eqnarray}\label{eq:S}
\hat S^{ij}{}_k &:= & {} - 2 Q \left( 2\delta_k{}^{(i}\hat{\cal C}^{j)}
        -\delta^{ij}\hat{\cal C}_{k} \right)
        - \alpha_0 \left(2 \delta^{(i}{}_k k^{j)}  - 
        h^{ij} k_k \right) \hat \lambda                          \nonumber \\
                                                       &    &
       {} + \sqrt{3}\alpha_0 k^m \left(2 \hat 
          \lambda^{ij}{}_{km} - h^{ij} h_{rs}
       \hat \lambda^{rs}{}_{km}\right) \; .
\end{eqnarray}
This system of equations naturally splits up in two subsystems, since
the equations (\ref{lambda0}--\ref{c1}) couple to the equations
(\ref{lambda3}--\ref{c3}) only via the ``source'' term in (\ref{c3}).
In the following, we will first establish that the solutions 
to the subsystem (\ref{lambda0}--\ref{c1}), and hence the coupling
term in (\ref{c3}), asymptotically decay to zero.  
In a second step, we consider this coupling as a given, decaying
source, and discuss the asymptotic behaviour of solutions to the
subsystem (\ref{lambda3}--\ref{c3}).
\begin{lemma}
\label{lemma1}
Let $\cal H$ be the space of the Fourier transformed 
$\hat\lambda^{ij}{}_{kl}\in L^2$, and let ${\cal D}\subset {\cal H}$ be 
the subspace defined by
$\hat \lambda^{ij}{}_{ks} k^{s} =0$. Then $\cal D$ is invariant under 
time evolution, and the trivial solution 
$\hat \lambda^{ij}{}_{kl}=0$ is asymptotically stable for 
the evolution restricted to $\cal D$. 
\end{lemma}
\textbf{Proof:}
Multiplying equation~(\ref{lambda4}) 
by $k_m$, antisymmetrizing, and using that $\hat C^{ij}{}_{[kl} k_{m]}=
\hat M^{ij}{}_{[k} k_l k_{m]}=0$, we obtain
\begin{equation}
  \label{eq:D}
  \dot{\hat \lambda}{}^{ij}{}_{[kl} k_{m]} 
= - \beta \,\hat \lambda^{ij}{}_{[kl} k_{m]}\;.
\end{equation}
Next we note that for a function 
$\hat\lambda^{ij}{}_{kl}$ in $\cal H$, the component 
$(\hat\lambda^{ij}{}_{kl})^\parallel$ in 
$\cal D$  
is given by $(\hat\lambda^{ij}{}_{kl})^\parallel = 
\hat \alpha^{ij} k^r \varepsilon{}_{rkl}$, where 
$\hat \alpha^{ij} = {\hat \lambda^{ij}{}_{[kl} k_{r]}
  \varepsilon^{klr}}/({6k^2})$.  
Equation (\ref{eq:D}) is, therefore, equivalent
to
\begin{equation}
\label{eq:Dparallel}
  (\dot{\hat \lambda}{}^{ij}{}_{kl})^\parallel 
= - \beta \,(\hat \lambda^{ij}{}_{kl})^\parallel\;,
\end{equation}
which proves lemma~\ref{lemma1}.

By direct inspection of the evolution equations, it follows that the
equation for the component of $\hat\lambda{}^{ij}{}_{kl}$ in 
the subspace $\cal D$ decouples.
It is, therefore, sufficient to concentrate on the evolution in the
space  $C\!F_\lambda\oplus C\!F_C$ which comprises those functions
$(\hat \lambda,\hat \lambda^i,\allowbreak \hat
\lambda^{ij}{}_{kl},\allowbreak \hat{\cal C}, \hat{\cal C}^i,\allowbreak 
\hat{\cal C}^{ij}{}_{kl})\in L^2$ for which 
$\hat\lambda{}^{ij}{}_{kl}\in {\cal D}^\perp$.
Here,  ${\cal D}^\perp$ denotes the $L^2$ complement of 
${\cal D}$ in ${\cal H}$, which, as easily seen, is spanned by the elements 
$\hat\lambda{}^{ij}{}_{kl}\in L^2$ satisfying 
$\hat \lambda{}^{ij}{}_{[kl} k_{m]}=0$.\footnote{For functions
  $\hat\lambda^{ij}{}_{kl}$ in $\cal D$ only the components along
  $k^m$ are non-trivial, $\hat\lambda^{ij}{}_{kl} = -{2}k_{[l}
  \hat\lambda^{ij}{}_{k]m} k^m/k^2$. This can be seen by solving $\hat
  \lambda^{ij}{}_{[kl} k_{m]} =0$, and by using the antisymmetry in
  the lower indices of $\hat\lambda^{ij}{}_{kl}$.}
Since for the constraint variable $\hat C {}^{ij}{}_{kl}$, the same
property is fulfilled, $\hat C{}^{ij}{}_{[kl} k_{m]}=0$, this shows
that the spaces $C\!F_\lambda$ and $C\!F_C$ are naturally isomorphic,
$C\!F_\lambda\approx C\!F_C=:C\!F$.

To simplify the notation, and to display the structure of the
evolution equations considered more transparently, let us introduce 
the following operator $\mbox{\boldmath$E$}$ acting 
on functions $\mbox{\boldmath{$v$}} \equiv (v,v^i,v^{ij}{}_{kl})$ in $C\!F$:

\begin{equation}\label{eq:E}
        \mbox{\boldmath$E\mbox{(}v\mbox{)}$} \equiv
        \left(E(\mbox{\boldmath$v$}),
         E^i(\mbox{\boldmath$v$}), E^{ij}{}_{kl}(\mbox{\boldmath$v$})
         \right)\;,
\end{equation} 
where
\begin{eqnarray}
        E(\mbox{\boldmath$v$})                  & \equiv &
         \sqrt{3} \alpha_0 v^{rl}{}_{rm} k^m k_l -
        2\alpha_0 v k^nk_n \;,                                    
\label{e}\\
        E^i(\mbox{\boldmath$v$})                & \equiv & 
        -\frac12 \alpha_1 \left(v^i
        k^nk_n + v^n k^ik_n\right) \;,                   \label{ei}\\
        E^{ij}{}_{kl}(\mbox{\boldmath$v$})      & \equiv & -
        \sqrt{3} \alpha_0 \left(v^{ij}{}_{kn} k^n k_l - v^{ij}{}_{ln}
          k^n k_k \right)
        \nonumber\\
                                                        &    &
        + \alpha_0 v \left(\delta_k{}^{(i} k^{j)} k_l
               -\delta_l{}^{(i} k^{j)} k_k \right)      \; .  \label{eijrs}
\end{eqnarray}
Taking advantage of these definitions, the evolution system
(\ref{lambda0}-\ref{c4}) restricted to the subspace $C\!F\oplus C\!F$ can
be rewritten as
\begin{equation}                                
\label{lambdaCsystem} 
\frac{d}{dt}\left( \begin{array}{c}
                            \mbox{\boldmath$\lambda$} \\ 
                            \mbox{\boldmath$C$}
            \end{array}
     \right)
=    \left( \begin{array}{cc}
           -\mbox{\boldmath $S$} &  \mbox{\boldmath $\Gamma$}  \\
            \mbox{\boldmath $E$} & i\mbox{\boldmath $A$}
            \end{array}
     \right)
     \left( \begin{array}{c}
                         \mbox{\boldmath$\lambda$} \\ 
                            \mbox{\boldmath$C$}   
            \end{array}
     \right)  
=: 
\mbox{\boldmath$P$}    
            \left( \begin{array}{c}
                         \mbox{\boldmath$\lambda$} \\ 
                            \mbox{\boldmath$C$}   
            \end{array}
     \right)\quad,
\end{equation}
where 
\mbox{\boldmath $S$} and \mbox{\boldmath $\Gamma$} are
diagonal matrices determined by the 
parameters $\beta$ and $\alpha_i$, respectively, and where 
$\mbox{\boldmath $A$}$ is an operator of the form 
$\mbox{\boldmath $A$}^m k_m$. 
 
In a next step, we show that the operator $e^{\mbox{\boldmath $P$}t}$ is 
bounded with respect to a suitably chosen norm.
To this end, we first establish the
following
\begin{lemma}
\label{lemma2}
The operator $H_{\lambda}:=-\mbox{\boldmath 
$\Gamma$}^{-1}H_c \mbox{\boldmath$E$}$
considered as a matrix-valued field on the Fourier space
$\mbox{\boldmath$R$}^3$
is symmetric and coercive with respect to the inner 
product $\langle \mbox{\boldmath$u$},
\mbox{\boldmath$v$}\rangle:=\bar{u}v+e_{ij}\bar{u}^iv^j+
e_{ip}e_{jq}e^{kr}e^{ls}\bar{u}^{ij}{}_{kl}v^{pq}{}_{rs}$. 
That is, $\langle \mbox{\boldmath$u$},H_{\lambda}
\mbox{\boldmath$v$}\rangle =\langle H_{\lambda}
\mbox{\boldmath$u$},\mbox{\boldmath$v$}\rangle $ for all
$\mbox{\boldmath$u$},\mbox{\boldmath$v$}\in C\!F$,
and there exists a constant $c>0$ such that 
$\langle \mbox{\boldmath$u$},H_{\lambda}\mbox{\boldmath$u$}\rangle  
\geq ck^2\langle \mbox{\boldmath$u$},\mbox{\boldmath$u$}\rangle$
for all $\mbox{\boldmath$u$}\in C\!F$.
\end{lemma}
\textbf{Proof:} 
We have
\begin{eqnarray} &&
 \langle \mbox{\boldmath$u$},\mbox{\boldmath$\Gamma$}^{-1}H_c
         \mbox{ \boldmath$E\mbox{(}v\mbox{)}$}
 \rangle 
  -
 \langle \mbox{\boldmath$\Gamma$}^{-1}H_c
         \mbox{ \boldmath$E\mbox{(}u\mbox{)}$}, \mbox{\boldmath$v$}
 \rangle                                        \nonumber\\
&& 
   = 
 \frac1{3\alpha_0}\bar{u}
       \left(-2\alpha_0 v k_nk^n 
        +\sqrt{3} \alpha_0 v^{kl}{}_{km}k^m k_l\right)          \nonumber \\
&& \hphantom{=}
   + \frac1{\alpha_1}\bar{u}^i
        \left(-\frac{\alpha_1}2
        \left(v^j k_nk^n + k^j v^lk_l\right)
                                 \right)e_{ij}  \nonumber \\
&& \hphantom{=}
   + \frac1{2 \sqrt{3} \alpha_0}e_{im}e_{jn}e^{kp}e^{lq}
        \bar{u}^{ij}{}_{kl}
        \left(-2 \sqrt{3} \alpha_0 v^{mn}{}_{ps}k^sk_q 
         +2\alpha_0 v \delta^m{}_p k^n k_q
                                         \right)      \nonumber\\
&& \hphantom{=}
   -\frac1{3\alpha_0} v 
        \left(-2\alpha_0 \bar{u} k_nk^n 
         +\sqrt{3} \alpha_0 \bar{u}^{kl}{}_{km}k^m k_l\right)   \nonumber \\
&& \hphantom{=}
   - \frac1{\alpha_1} v^i
        \left(-\frac{\alpha_1}2
        (\bar{u}^j k_nk^n + k^j\bar{u}^lk_l)
                                \right)e_{ij}   \nonumber \\
&& \hphantom{=}
   +\frac1{2 \sqrt{3} \alpha_0}e_{im}e_{jn}e^{kp}e^{lq}
        v^{ij}{}_{kl}
        \left(-2 \sqrt{3} \alpha_0 \bar{u}^{mn}{}_{ps}k^sk_q 
         +2\alpha_0 \bar{u} \delta^m{}_p k^n k_q
                                              \right) \nonumber \\
&& 
   = 0. \nonumber
\end{eqnarray}
The remaining part of the proof is given in appendix~\ref{ap:coercivity}, 
where we show that $H_\lambda$ is coercitive with constant $c=1/4$. 

With the help of lemma~\ref{lemma2}, it is now easy to prove
\begin{lemma}
\label{lemma3}
The matrix-valued fields $\mbox{\boldmath$P_\pm$}$,
\begin{equation}
  \label{eq:defP_0}
 \mbox{\boldmath$P_+$}
:=
 \left( 
    \begin{array}{ll}
  \mbox{\boldmath$S$} & 0 \\
    0 & 0 \\
    \end{array}
  \right)\;,\quad
\mbox{\boldmath$P_-$}:= \left( 
    \begin{array}{ll}
                    0  &  \mbox{\boldmath$\Gamma$} \\
  \mbox{\boldmath$E$}  & i\mbox{\boldmath $A$}     \\
    \end{array}
  \right)\;\;,
\end{equation}
are hermitian respectively anti-hermitian with respect to 
the inner product
\begin{equation}
  \label{eq:HT}
   \langle (\mbox{\boldmath $\lambda_1$},\mbox{\boldmath $C_1$}),
    H_T (\mbox{\boldmath $\lambda_2$},\mbox{\boldmath $C_2$})
   \rangle 
\,:= \,
    \langle \mbox{\boldmath $\lambda_1$}, H_{\lambda}
           \mbox{\boldmath$\lambda_2$} 
   \rangle 
+
   \langle \mbox{\boldmath $C_1$},H_c \mbox{\boldmath $C_2$}
   \rangle\;\,.  
\end{equation}
\end{lemma}
\textbf{Proof:} Since $\mbox{\boldmath$S$}=\beta\mbox{\boldmath$I$}$, 
the statement for $\mbox{\boldmath$P_+$}$ is trivially true.
The anti-symmetry of 
$\mbox{\boldmath$P_-$}$ follows directly from lemma~\ref{lemma2},
and the symmetry of $\mbox{\boldmath $A$}$ with respect to $H_c$.

Taking advantage of lemma~\ref{lemma3}, we now obtain the following important
estimate for the operator $\mbox{\boldmath$P$}=\mbox{\boldmath$P_+$}+
\mbox{\boldmath$P_-$}$:
\begin{equation}
\label{eq:HP}
      H_T \mbox{\boldmath $P$} 
    + \mbox{\boldmath $P$}^{\dagger} H_T 
  =       
      H_{\lambda}\mbox{\boldmath $S$}
    + \mbox{\boldmath $S$} H_{\lambda}
  = 
    - 2 \beta H_{\lambda} 
  \leq 
     -2 \beta \frac14k_nk^n 
   \leq 0\;,
\end{equation}
where, for any Hermitian matrix \mbox{\boldmath$M$}, the inequality
$\mbox{\boldmath$M$}\leq 0$ means 
$\langle \mbox{\boldmath$v$},\mbox{\boldmath$Mv$}\rangle \leq 0$ for all
\mbox{\boldmath$v$}.\footnote{Clearly, there are other possible choices
  of the operators $\mbox{\boldmath $S$}$ which lead to the same
  inequality. Here we have restricted to the simplest possibility, but
  for practical applications, alternative choices might be better suited.}

The symmetry and coercivity of the operator $H_\lambda$ imply that
$H_\lambda$ can be used to define a scalar product on a (dense) subspace 
${\cal D}(C\!F)$ of the Hilbert space $C\!F$,\footnote{In physical space, 
the relevant function
   space equipped with the norm corresponding to
   the above scalar product is very similar to the Sobolev 
   space $H^1_0$.} 
which, in turn, shows that the operator $H_T=H_c+H_\lambda$ 
gives rises to a scalar product on $C\!F\oplus {\cal D}(C\!F)$.

As is well known (see, for instance,~\cite{KrL89IV}), the estimate 
(\ref{eq:HP}) implies that for all $t>0$, the operator 
$e^{\mbox{\boldmath $P$}t}$ is 
bounded with respect to the norm defined by $H_T$. Hence, the 
initial value
problem for the system considered is well posed. Moreover,
all solutions with initial data which are bounded with respect to this 
norm remain bounded for all positive times. Thus Laplace transformation
techniques can be applied~\cite{KrL89IV},  
and the relevant questions are the {\em sign\/} of the real part of 
the eigenvalues 
of $\mbox{\boldmath$P$}$, and how fast they approach zero as 
the wave number $k=\sqrt{k^ik_i}$ goes to zero. 
Hence, the proof is reduced to the eigenvalue problem for the operator
\mbox{\boldmath $P$},
\begin{equation}\label{peigen}
        \mbox{\boldmath $P$} \left( \begin{array}{c}
                           \mbox{\boldmath $\lambda_s$}   \\ 
                           \mbox{\boldmath $C_s$} 
            \end{array}
     \right)
   =
        s 
        \left( \begin{array}{c}
                           \mbox{\boldmath $\lambda_s$}   \\ 
                           \mbox{\boldmath $C_s$} 
            \end{array}
     \right)\;.
\end{equation}

Then we have the following
\begin{lemma}
\label{lemma4}
The eigenvalues of the above system have non-positive real part and
furthermore there exist  positive constants $c_1$ and $w_1$ 
such that
\begin{equation}
  \label{eq:Resbound}
  \Re(s) \leq -c_1\frac{k^2}{w_1+k^2} 
\end{equation}
for all wave vectors $k_i$
\end{lemma}
\textbf{Proof:}
{}From the {\boldmath $\lambda$}-rows of the eigenvalue equation, we get
\begin{eqnarray}
  \label{eq:Coflambda}
  \mbox{\boldmath $C_s$} = 
  (s+\beta)\mbox{\boldmath $\Gamma$}^{-1}\mbox{\boldmath $\lambda_s$}
  \; .
\end{eqnarray}
Using this in the {\boldmath $C$}-rows, we next obtain
\begin{equation}
  \label{eq:eigenlambda}
  \left(\mbox{\boldmath $E$} + (s+\beta)(-s\mbox{\boldmath $I$} 
    + i\mbox{\boldmath $A$})
                         \mbox{\boldmath
                           $\Gamma$}^{-1}\right)\mbox{\boldmath $\lambda_s$}
                         = 0 \; .
\end{equation}
Multiplying the above equation by 
$-(\mbox{\boldmath $\Gamma$}^{-1})^{\dagger}\mbox{\boldmath $H_c$}$
from the left and subsequently contracting with {\boldmath
  $\lambda_s$}, we find the following second order equation for the
eigenvalue $s$:
\begin{equation}
  \label{eq:second-order}
  \langle 
    \mbox{\boldmath $\lambda_s$},H_{\lambda}\mbox{\boldmath $\lambda_s$}\rangle  + 
    (s+\beta)\left(s\langle \mbox{\boldmath $\lambda_s$},
    (\mbox{\boldmath $\Gamma$}^{-1})^{\dagger}
    H_c\mbox{\boldmath $\Gamma$}^{-1}\mbox{\boldmath $\lambda_s$}\rangle  
    - i\langle \mbox{\boldmath $\lambda_s$},
    (\mbox{\boldmath $\Gamma$}^{-1})^{\dagger}H_c\mbox{\boldmath $A$}
    \mbox{\boldmath $\Gamma$}^{-1}
    \mbox{\boldmath $\lambda_s$}\rangle \right) = 0 \; .
\end{equation}
The established properties of the involved operators imply that
\begin{equation}
 c(k_i^0)k^2 := 
\frac{\langle \mbox{\boldmath $\lambda_s$},
  H_{\lambda}\mbox{\boldmath $\lambda_s$}\rangle }
     {\langle \mbox{\boldmath $\lambda_s$},
        (\mbox{\boldmath $\Gamma$}^{-1})^{\dagger}H_c\mbox{\boldmath $\Gamma$}^{-1}
       \mbox{\boldmath $\lambda_s$}\rangle }
\end{equation} 
is positive for $k_i\neq 0$, and that
\begin{equation}
  b(k_i^0)k := 
    \frac{\langle \mbox{\boldmath $\lambda_s$},
        (\mbox{\boldmath $\Gamma$}^{-1})^{\dagger}H_c\mbox{\boldmath
          $A$}
        \mbox{\boldmath $\Gamma$}^{-1}
       \mbox{\boldmath $\lambda_s$}\rangle }
     {\langle \mbox{\boldmath $\lambda_s$},
        (\mbox{\boldmath $\Gamma$}^{-1})^{\dagger}H_c\mbox{\boldmath $\Gamma$}^{-1}
      \mbox{\boldmath $\lambda_s$}\rangle }
\end{equation} 
is real, where $k_i^0$ denotes the unit vector in the direction of 
$k_i$, and $k$ is the norm of $k_i$. 
Thus we have for each direction of $k^i$
\begin{equation}
  \label{eq:second-order-dir}
  (s+\beta)(s-ibk) + ck^2 =0
\end{equation}
with $\beta$, $b$, $c$ real and $\beta$, $c$ positive. 
For this equation we prove in appendix~\ref{AnhangII} that the real
part of the roots satisfies the desired inequality, which establihes
the result for each direction of the wave vector $k_i$.  
Using the maximal values of $-c_1\frac{k^2}{w_1+k^2}$ on the 2-sphere
of directions of $k_i$, we obtain the final inequality.

With this bound on the decay constants, it is now easy to prove
asymptotic stability for the subsystem~(\ref{lambda0}--\ref{c4}). 
Splitting the set of solutions in a part with frequencies with $k<1$, and
another with $k\geq 1$, the above bound tells us that the solutions of
the higher frequency part decay faster than $e^{-\frac{c_1
    t}{w_1+1}}$, while the decay of the solutions of the low
frequency part can be estimated as in \cite[lemma 1 and 2 of section
III]{KreissG:StabCL}.

We now turn attention to the second set of equations, given by
(\ref{lambda3}) and (\ref{c3}), and establish the following
\begin{lemma}
\label{lemma5}
Let ${\cal H}_3$ be the space of the Fourier transformed
$(\hat\lambda^{ij}{}_k,\hat C^{ij}{}_k)\in L^2$. Then ${\cal H}_3$ is invariant
under time evolution, and the trivial solution 
$(\hat\lambda^{ij}{}_k,\hat C^{ij}{}_k)=0$ is asymptotically stable
for the evolution restricted to ${\cal H}_3$. 
\end{lemma}
\textbf{Proof:} 
In a first step, we discuss the equation for the component of 
a solution in the 
subspace 
\begin{equation}
{\cal D}_{3}:=\{\,(\hat\lambda^{ij}{}_k,\hat C^{ij}{}_k)\in L^2 
\,\mid\,
                   \hat \lambda^{ij}{}_m k^m=\hat C^{ij}{}_m 
                   k^m=0 \,\}\;.                   
\end{equation}
Taking advantage of equation (\ref{M}), (\ref{cijkl}), and (\ref{lambda3}),
we obtain
\begin{eqnarray}
  \label{eq:D2perp}
  \dot{\hat{\lambda}}{}^{ij}{}_{[k}\,k_{l\,]} 
    &=&- \beta\,\hat{\lambda}^{ij}{}_{[k}\,k_{l\,]} 
    +\alpha_3\,
\hat{C}^{ij}{}_{[k}\,k_{l\,]}\;,\\
i\,\hat{C}^{ij}{}_{[k}\,k_{l\,]}&=&\hat{C}^{ij}{}_{kl}-\frac{1}{2}
\delta^{ij}\delta_{mn}\hat C^{mn}{}_{kl}\label{eq:D2p2}\;,
\end{eqnarray}
which implies that the space ${\cal D}_3$ is invariant under time evolution.
As already shown, the constraint variable $\hat{C}^{ij}{}_{kl}$ 
asymptotically decays to zero. The dynamics in ${\cal D}_3$ is, therefore,
described by a system of ordinary differential equations of the
form
$\dot{u} = -u + f$, where $f$ is a given source with $f \to 0$ as 
$t \to \infty$. Since
any solution to this system satisfies $u \to 0$ as 
$t \to \infty$,\footnote{For a proof, choose $T$ such that $f(t) <
  \eps/2$ for all $t > T$. Since the general solution to the above
  system is given by $u(t) = e^{-t}(u(0)+ \int_0^t e^{\tilde t} 
  f(\tilde t) d\tilde t)$, it follows that $u(t) \leq e^{-t}(u(0)+ 
  \int_0^T e^{\tilde t} f(\tilde t) d\tilde t - \eps e^T/2) + \eps/2$.
  Hence, for a sufficiently large time $t_0>T$, the absolute value of
  the first term becomes smaller than $\eps/2$, which implies
  $|\,u(t)\,|<\eps$ for all $t>t_0$.}
it follows that solutions in ${\cal D}_3$ decay with time.
  
It remains to discuss the complementary subspace 
${\cal D}_3^\perp$,
\begin{equation}
{\cal D}_3^\perp=\{\,(\hat\lambda^{ij}{}_k, \hat{C}^{ij}{}_k) 
\in L^2 \,\mid\,
\hat \lambda^{ij}{}_{[k}\,k_{l\,]}= 
\hat C^{ij}{}_{[k}\,k_{l\,]}= 0 \,\}\;.
\end{equation}
For the component of a solution in this subspace, we find 
\begin{equation}                                
\label{lC2} 
\frac{d}{dt}\left( \begin{array}{c}
                            \mbox{\boldmath$\lambda_3$} \\ 
                            \mbox{\boldmath$C_3$}
            \end{array}
     \right)
\;=\;    \left( \begin{array}{cc}
           -\beta & \alpha_3  \\
           -\alpha_3\,k^2   &ik_mN^m 
            \end{array}
     \right)
     \left( \begin{array}{c}
                         \mbox{\boldmath$\lambda_3$} \\ 
                            \mbox{\boldmath$C_3$}   
            \end{array}
     \right)
\;\;-\;
\left( \begin{array}{c}
                            0 \\ 
                            F^{ij}k_k
            \end{array}
     \right)\;\;,
\end{equation}
where 
$(\mbox{\boldmath$\lambda_3$},\mbox{\boldmath$C_3$}):=
(\hat\lambda^{ij}{}_k,\hat C^{ij}{}_k)^\perp\in {\cal D}_3^\perp$,
and where
$\hat F^{ij}k_k$ is a shorthand for the perpendicular component of 
the source term $\hat S^{ij}{}_k$, 
$
\hat F^{ij}=\hat S^{ij}{}_m k^m/k^2\;.
$
Thus, as expected, the subspace ${\cal D}_3$ is invariant as well.

Since $\hat P^{ij}$ and consequently $\hat C^i/|k|=\hat P^{im}k_m/|k|$ 
are contained in $L^2$, equation (\ref{eq:S}) implies that 
the same is true for $\hat F^{ij}$, $\hat F^{ij}\in L^2$.
Furthermore, the real part of the eigenvalues of the system 
(\ref{lC2}) can, as in lemma~\ref{lemma4}, 
be estimated by the inequality (\ref{eq:Resbound}), 
albeit for different constants. 
Adopting a similar reasoning as in the previous discussion, and
applying lemma~1 and 2 of \cite{KreissG:StabCL} 
to this system, it follows that solutions in ${\cal D}_3^\perp$ 
also decay with time. 

This completes the proof of lemma~\ref{lemma5} and hence the proof of
our main result.

\section{Conclusions}
\label{sec:conclu}

In the present paper we have shown that an arbitrary system of
symmetric hyperbolic evolution equations with constraints admits extensions
to symmetric hyperbolic systems which reproduce the original dynamics on the
embedded constraint submanifold. We have given analytical evidence that the
class of extensions proposed is sufficiently rich to contain systems
for which the embedded constraint submanifold is an attractor of the time
evolution. For the Einstein equations, we have constructed an extended 
evolution system for which, at least in the linearized case, 
this property is fulfilled. 

It is natural to expect that, by making use of techniques developed
in~\cite{KreissG:StabCL}, the results proven for the linearized
Einstein equations can be generalized to the regime of non-linear
general relativity describing space-times in the vicinity of Minkowski
space. 
However, to establish similar results for more extended regions of the
phase space of general relativity is well beyond the scope of present
analytic techniques.

Numerical experiences with the Navier-Stokes equations for
incompressible fluids show that asymptotic stability of the constraint 
submanifold is essential for accurate results~\cite{Kr98pC}.
For this system, techniques with a very similar effect have been used to
include the incompressibility constraint into the evolution equations.
On the basis of this observation, and the results established for
linearized gravity, we suspect that the extensions of Einstein's
equations constructed could be of interest when obtaining numerical
solutions to general relativity.  
Numerical experiments testing aspects of this conjecture are in progress.

\subsection*{Acknowledgments}
OB would like to thank Chris Beetle for discussions and interesting
comments.
SF and OR gratefully acknowledges the hospitality of the Albert Einstein
Institut, Potsdam, where part of this work was carried out.
${\cal P\!\! H}$ would like to thank Bernd Schmidt for interesting
discussions.

\appendix

\section{Proof of coercivity}
\label{ap:coercivity}

In this appendix we show that 
\begin{equation}\label{unitnegE}
\langle \mbox{\boldmath$u$},H_\lambda\mbox{\boldmath$u$}\rangle  \;
   \geq
         \frac14 k_nk^n
\end{equation}
for unitary $\mbox{\boldmath$u$}$ satisfying 
$u^{ij}{}_{rs}=u^{(ij)}{}_{[rs]}$ and 
 $u^{ij}{}_{[rs}k_{l]}=0$, as needed for lemma~\ref{lemma2}. 
We treat this as the problem of extremizing the quadratic function
of $\mbox{\boldmath$u$}$ on the left-hand side of (\ref{unitnegE})
under the constraint condition $\langle
\mbox{\boldmath$u$},\mbox{\boldmath$u$}\rangle =1$.

{}From (\ref{e}--\ref{eijrs}), we obtain
\begin{eqnarray} &&
 \langle \mbox{\boldmath$u$},H_\lambda\mbox{\boldmath$u$}\rangle 
 +\tau k^nk_n \left(1-\langle\mbox{\boldmath$u$},\mbox{\boldmath$u$}\rangle \right) 
 =                                               \nonumber\\
                                                &&
        \frac23 k^nk_n u\bar{u}
        +\frac12 k^nk_nu^i\bar{u}_i
        +\frac12 u^ik_i\bar{u}_nk^n
        +        u^{ij}{}_{rs}k^s\bar{u}_{ij}{}^{rm}k_m
        -\frac{\sqrt{3}}{3}\bar{u}u^{ij}{}_{is}k^sk_j
                                                \nonumber\\
                                                &&
        -\frac{1}{\sqrt{3}}u\bar{u}_{ij}{}^{is}k_sk^j
        +\tau k^nk_n 
        \left( 1
               - u\bar{u}
               -u^i\bar{u}_i
               -u^{ij}{}_{rs}\bar{u}_{ij}{}^{rs} 
        \right) \; ,
\end{eqnarray}
where $\tau$ is a Lagrange multiplier and where indices are raised
and lowered with $e_{ij}$.
To simplify the algebra, we choose a basis in which $k^n=(0,0,k)$. 
Then $u^{ij}{}_{rs}=0$, except when $s=3$. 
Hence, 
\begin{eqnarray} &&
        F(\mbox{\boldmath$u$},\tau) 
\equiv
 \langle \mbox{\boldmath$u$},H_\lambda\mbox{\boldmath$u$}\rangle 
+\tau k^nk_n (1-\langle \mbox{\boldmath$u$},\mbox{\boldmath$u$}\rangle)  
=                                               \nonumber\\
                                                &&
        k^2\left(
        \frac23  u\bar{u}
        +\frac12 u^i\bar{u}_i
        +\frac12 u^3\bar{u}_3
        +        u^{ij}{}_{r3}\bar{u}_{ij}{}^{r3}
        -\frac{1}{\sqrt{3}} \bar{u}u^{i3}{}_{i3}\right.  \nonumber\\
                                                &&
        \left.
        -\frac{1}{\sqrt{3}} u\bar{u}_{i3}{}^{i3}
        +\tau 
        \left(
                1
                -u\bar{u}
                -u^i\bar{u}_i
                -u^{ij}{}_{rs}\bar{u}_{ij}{}^{rs}
        \right)
        \right) \; .
\end{eqnarray}

The function $F(\mbox{\boldmath$u$},\tau)$ is extremized at points
$(\mbox{\boldmath$u$},\tau)$ where 
\begin{eqnarray}
         \frac{\partial F}{\partial \mbox{\boldmath$u$}}
        +\frac{\partial F}{\partial \mbox{\boldmath$\bar{u}$}}
  & =& 0 \; ,                                       \label{relinearhom}\\
         \frac{\partial F}{\partial \mbox{\boldmath$u$}}
        -\frac{\partial F}{\partial \mbox{\boldmath$\bar{u}$}}
  & =& 0 \; ,                                       \label{imlinearhom}\\
        \frac{\partial F}{\partial \tau}
  & =&
        0 \; .                                      \label{condition}
\end{eqnarray}
Equation (\ref{condition}) is the requirement that \mbox{\boldmath$u$} has
unit length.  
Equation (\ref{relinearhom}) and (\ref{imlinearhom}) constitute a
homogeneous linear system of equations for the real and imaginary parts
of \mbox{\boldmath$u$}.  
Since  \mbox{\boldmath$u$} cannot vanish, the determinant of the
linear system has to vanish.
Up to numerical factors, this is given by
\begin{equation}
        \left(2\tau-1\right)^{13}
        \left(\tau-1\right)^2
        \left(\tau-\frac16\right) \; .
\end{equation}
As easily verified, $\tau=1$ yields the following minimal value of
$F(\mbox{\boldmath$u$},\tau)$ (when evaluated at unit
\mbox{\boldmath$u$} such that (\ref{relinearhom})--(\ref{imlinearhom})
are satisfied):
\begin{equation}
        F(\mbox{\boldmath$u$}_{min},1) = \frac14 k^2 \; ,
\end{equation}
from which (\ref{unitnegE}) follows.  The other extreme values of
$F(\mbox{\boldmath$u$},\tau)$ are $(5/3)k^2$ and $(1/2)k^2$ for
$\tau=1/6,1/2$.


\section{On the proof of lemma~\ref{lemma4}}
\label{AnhangII}

In this appendix we prove that the roots $s_\pm$ of the
polynomial 
\begin{equation}
P(s)=s^2+ s(\beta-ibk)+ck^2\;
\end{equation}
are subject to the inequality
\begin{equation}
  \label{eq:rootineq}
  \Re (s_\pm) \leq -c_1\frac{k^2}{w_1+k^2}\;, 
\end{equation}
where $c_1 = \beta c/(b^2+4c)$ and $w_1={\beta^2}/(b^2+4c)$. 
As in the body of the text, it is assumed that the parameters
of $P$ are real, and that $\beta$ and $c$ are strictly positive. 

To begin with, let us rewrite the polynomial $P$, and the above estimate
in terms of suitably rescaled parameter. Defining
\begin{equation}
\tilde{s}=s/\beta\;,\quad \tilde{k} = 2k\sqrt{b^2+4c}\,/\beta\;,\quad 
\tilde{b}=b/\sqrt{b^2+4c}\;\;,
\end{equation}
and dropping tildes, we obtain for the polynomial
\begin{equation}
\label{scaledP}
P/\beta^2=s^2 + s(1-ibk)+ (1-b^2)k^2/4\;.
\end{equation}
The estimate for the roots in terms of the 
scaled parameters assumes the form
\begin{equation}
   \Re (s_\pm) \leq -\frac{\gamma^2}{4}\frac{k^2}{1+k^2}\,,
\end{equation}
where $\gamma^2:=1-b^2\in (0,1]\,$. 

As easily verified, the roots
of the scaled polynomial satisfy
\begin{equation}
  \label{eq:roots}
  \max\{\,\Re(\,2s_+)\,,\,\Re(\,2s_-)\,\} = -1 +  
    |\,\Re\,\sqrt{1-k^2 +2ibk}\,|\,\;.
\end{equation} 
It is, therefore, sufficient to show that
\begin{equation}
  \label{eq:ineq1}
  |\,\Re\,\sqrt{1-k^2+2ibk}\;|\leq 1-\frac{\gamma^2}{2}\frac{k^2}{1+k^2}\;\,.
\end{equation} 

To give a proof of this inequality, we first evaluate the identity
\begin{equation}
2\,|\,\Re\,\sqrt{z}\,|^{\,2}=|\,z\,|+\Re(z)\
\end{equation}
for $z:={1-k^2+2ibk}\;$,
\begin{eqnarray}
  2\,|\,\Re\,\sqrt{z}\,|^{\,2}&=&
\sqrt{(1-k^2)^2+4b^2k^2}+(1-k^2) \nonumber \\
&=&\sqrt{(1+k^2)^2-4\gamma^2 k^2}-(1+k^2)+2 \;.\nonumber
\end{eqnarray}
Hence,
\begin{eqnarray}
  \label{eq:estimate}
  \,|\,\Re\,\sqrt{z}\,|^{\,2}
&=& 1+(1+k^2)\left\{\sqrt{1-{4\gamma^2 k^2}/{(1+k^2)^2}} - 1\right\}/\,2 
\nonumber \\
&\leq &1 -\gamma^2\frac{k^2}{1+k^2}\;,
\end{eqnarray}
where we have used the estimate $\sqrt{1-x} \leq 1- x/2$, which holds for 
$x\leq 1$.
Taking advantage of the latter estimate once again, it follows that 
\begin{equation}
  \label{eq:estimate2}
  |\,\Re\,\sqrt{z}\,|\,\leq \,1-\frac{\gamma^2}{2}\frac{k^2}{1+k^2}\;,
\end{equation}
which completes the proof of our claim.

\bibliography{/afs/aei-potsdam.mpg.de/u/reula/documentos/reula,/afs/aei-potsdam.mpg.de/u/pth/Biblio/biblio}
\bibliographystyle{alpha}

\end{document}